\documentclass[10pt,conference]{IEEEtran}
\IEEEoverridecommandlockouts
\usepackage{cite}
\usepackage{amsmath,amssymb,amsfonts}
\usepackage{algorithmic}
\usepackage{graphicx}
\usepackage{textcomp}
\usepackage{xcolor}
\usepackage{verbatim}
\usepackage{booktabs}
\usepackage{tabls}
\usepackage{url}
\usepackage{listings}
\usepackage{balance}
\usepackage{listings}
\usepackage{fancyvrb}
\usepackage[T1]{fontenc}
\usepackage[utf8]{inputenc}
\usepackage{hyperref}
\usepackage{url}
\usepackage{tikz,lipsum,lmodern}
\usepackage[most]{tcolorbox}
\usepackage{balance}
\usepackage{lastpage}
\usepackage{float}
\usepackage{balance}
\usepackage{lastpage}

\usepackage{amsmath}
\begin{document}

%\title{What are the Sentiments of Developers and End-Users in Game Engine Repositories? - An Exploratory Study}
\title{A Catalogue of Game-Specific Software Nuggets}
% \titlenote{An earlier version of the paper is available as a pre-print on \url{https://arxiv.org}, but we remove the citation to maintain the anonymity of submission.}
\author{
\IEEEauthorblockN{Vartika Agrahari, Sridhar Chimalakonda}
\IEEEauthorblockA{Research in Intelligent Software \& Human Analytics (RISHA) Lab\\Indian Institute of Technology Tirupati India \\
cs18m016@iittp.ac.in, ch@iittp.ac.in}
}

\maketitle
%theory, 
\begin{abstract}
With the ever-increasing use of games, game developers are expected to write efficient code, supporting several aspects such as security, maintainability, and performance. However, the continuous need to update the features of games in shorter duration might compel the developers to use anti-patterns, code smells and quick-fix solutions that may affect the functional and non-functional requirements of the game. These bad practices %are often termed as \textit{Anti-patterns}, which 
may lead to technical debt, poor program comprehension, and can cause several issues during software maintenance. Here, in this paper, we introduce \textit{``Software Nuggets''} as a concept that affects software quality in a negative way and as a superset of anti-patterns, code smells, bugs and software bad practices. We call these \textit{Software Nuggets} as \textit{``G-Nuggets''} in the context of games. While there exists empirical research on games, we are not aware of any work on understanding and cataloguing these \textit{G-Nuggets}. Thus, we propose a catalogue of \textit{G-Nuggets} by mining and analyzing 892 \textit{commits}, 189 \textit{issues}, and 104 \textit{pull requests} from 100 open-source GitHub game repositories. We use regular expressions and thematic analysis on this dataset for cataloguing game-specific \textit{Software Nuggets}. %We applied \textit{LDA} (Latent Dirichlet Allocation) to validate further and refine the categories. 
We present a catalogue of ten \textit{G-Nuggets} and provide examples for them available online at: \textit{\url{https://phoebs88.github.io/A-Catalogue-of-Game-Specific-Software-Nuggets}}. We believe this catalogue might be helpful for researchers for further empirical research in the domain of games and for game developers to improve quality of games.

\end{abstract}

\section{Introduction}
%Games are expanding their scope across the people of various ages and different sections of society.
The game industry continues to see expanding growth in terms of revenue and worldwide usage.
In a recent report, Newzoo predicts that 2021 will have revenue of about \$189.3 billion for mobile, PC, and console gaming\footnote{\url{https://newzoo.com/insights/articles/newzoos-games-trends-to-watch-in-2021/}}. 
%These statistics shows that the game industry is expanding with great pace.
However, increasing market demand for games and their widespread usage has posed a challenge for game developers to provide better quality games in less time \cite{kanode2009software}. This pressing need may sometimes force developers to use quick-fix solutions and can result in the violation of functional and non-functional requirements \cite{brown1998antipatterns, fowler1997refactoring}. Developers often tend to violate good design choices and coding practices, and may end up with bad practices or anti-patterns or code smells in software \cite{brown1998antipatterns, fowler1997refactoring}. Although these choices and practices may not directly affect the functionality of the program or cause any error, they may lead to long term future problems of maintainability, security, performance, and so on, and increasing technical debt.
Existing literature consists of several research studies regarding anti-patterns and code smells \cite{palomba2014anti,fard2013jsnose}. Brown et al. \cite{brown1998antipatterns} explain about anti-patterns stating that they are like misfits to a problem that should be prevented and avoided to lead to a better solution. However, these bad practices are often termed as anti-patterns, code smells, bad software practices, software bugs and similar concepts in different contexts \cite{palomba2014anti,fard2013jsnose} making it hard for developers to consider each of these concepts during development \cite{tahir2018can, taibi2017developers}. It is here, we believe a unified terminology and catalogue might help developers to better comprehend these varying definitions such as anti-patterns, code smells, bad software practices and software bugs.
%we look forward to propose a more holistic and integrated terminology which aggregates all of these. 
Thus, we propose \textit{"Software Nuggets"}\footnote{We call so, because the term \textit{Nuggets} in food domain is often considered as tasty and appetizing in the short term, but is often unhealthy in the long-term. Analogous to this, we propose \textit{Software Nuggets} as unhealthy for the quality of software systems.} as a generic concept that affects software quality in a negative way. 

\begin{tcolorbox}[colback=gray!5!white,colframe=gray!5!black]
\textbf{Software Nuggets:} We define \textit{Software Nuggets} as a concept that negatively affects quality of software and consider it as a superset of code smells, anti-patterns, bugs, and software bad practices. 

\textbf{G-Nuggets}: We refer the \textit{Software Nuggets} in the domain of games as \textit{G-Nuggets}.
\end{tcolorbox}

%\cite{appleton1997patterns} states that \textit{``If a pattern represents a `best practice', then an anti-pattern represents a `lesson learned' "}\footnote{Based on the definition by \cite{appleton1997patterns}, we use the term \textit{bad practice} and \textit{anti-pattern} interchangeably in the entire paper because bad practices are antonym to \say{best practices}, and \say{best practice} correspond to the desired pattern in software. Thus bad practices can be referred to as anti-pattern also.}. 

%about anti-patterns
%Existing literature consists of several research studies regarding anti-patterns and code smells \cite{palomba2014anti,fard2013jsnose}. \cite{brown1998antipatterns} explains about anti-patterns stating that they are like misfits to a problem that should be prevented and avoided to lead to a better solution. 
Despite the existence of a number of studies focusing on anti-patterns and code smells \cite{silva2016we, palomba2014anti, fowler1997refactoring}, understanding and analyzing the presence of these anti patterns and smells in games is still largely unexplored in the literature \cite{kanode2009software}. %motivating the need for our work. 
Games need to be considered as a peculiar domain over other software as it involves AI simulations, camera movements, actions of players, game mechanics, particle effects, and so on \cite{murphy2014cowboys}. They deal with real-time constraints and continuous rendering process, and, thus, the presence of \textit{Software Nuggets} in games may lead to degradation of game quality. Existing studies focused on psychological and social \cite{lee2006we}, educational \cite{giessen2015serious} and behavioural \cite{anderson2001effects} effects of games, but there is quite limited research on the software quality and bad practices in game development \cite{nystrom2014game,bjork2006games,borrelli2020detecting}, motivating the need for our work. 
% or discusses the bad practices in them %\cite{nystrom2014game,bjork2006games,borrelli2020detecting} ,. 

GitHub hosts a large number of software repositories including open-source games \cite{kalliamvakou2014promises}.
%, which is the focus of this paper. %Rapidly growing data on GitHub concedes the possibility to extract patterned information that can be utilized for better development of software in the future.
The increasing availability of data such as \textit{commits, issues} and \textit{pull requests} on GitHub can be leveraged to understand the problems faced by users and developers and corresponding potential solutions. Researchers have analyzed these artifacts of GitHub to understand the sentiments of developers \cite{venigalla2021sentiments}, their primary areas of interest by applying topic summarization \cite{sharma2017cataloging}, and similarity between projects \cite{zhang2017detecting}. In comparison to user feedback and surveys, they provide better and detailed information about the troubles faced by end-users and how they affect the usability and performance of a software \cite{kalliamvakou2014promises}. Thus, we considered the text corpus of \textit{commits}, \textit{issues}, and \textit{pull requests} to find the existence of \textit{Software Nuggets} in these these artifacts. % might contain information related to
% as they contain the problems faced by users and developers during the practical usage of games in the form of 
%anti-patterns, code smells, software bugs, and issues. 
%We believe that this text data include the problems which are closer to users and developers and thus hails importance.
% \begin{tcolorbox}[colback=grey!5!white,colframe=grey!5!black]
% Terminology description
% \end{tcolorbox}
Literature consists various catalogues on varied domains in software engineering such as android \cite{carvalho2019empirical}, energy patterns \cite{cruz2019catalog}, architecture \cite{garcia2009toward}, software metrics \cite{dalla2020towards} and many more which possibly emphasizes on the fact that domain-specific catalogues are more logical and implied.
Thus, considering game as a peculiar domain, there is need to have a specialized catalogue for bad practices in games.
In this paper, we propose a catalogue of game-specific \textit{Software Nuggets} that could serve as a checklist for game developers during the game development process. %Cataloguing could help in prioritising the development process and resources. 
%  Although, there are several catalogues gathering code smells and anti-patterns in different contexts in the literature \cite{cruz2019catalog,carvalho2019empirical}, to the best of our knowledge, there does not exist any catalogue in the context of games. 
Language-specific anti-patterns and detection techniques may not be sufficient to handle bad practices in games \cite{fard2013jsnose}. Unlike the existing literature which primarily focused on source code and other artifacts, we wish to leverage the potential of the text corpus of \textit{commits}, \textit{issues}, and \textit{pull requests} to answer the research question:  

\textbf{What are the most prevalent \textit{Software Nuggets} in the context of games?}
%and produce a catalogue of game-specific \textit{Software Nuggets}.

We apply a systematic methodology comprising of data collection followed by the commonly used thematic analysis \cite{fereday2006demonstrating} on 1185 text data records of 100 open-source games to propose a catalogue of ten \textit{G-Nuggets}. 
%Thematic Analysis is a well-known methodology used for observing and analysing the qualitative data . 
%We perform thematic analysis . 
We describe the methodology in detail in Section \ref{catalogue} and document the catalogue on our website as a reference for game developers. 
% \textbf{Q2: What are the distribution and occurrences of anti-patterns in games?}\newline
% To answer this research question, we try to find the co-occurrences of anti-patterns from the result of the thematic analysis. 
The contributions of the paper are as follows:
\begin{itemize}
    \item A catalogue of ten game-specific \textit{Software Nuggets} with a detailed discussion on each \textit{Nugget}. The detailed description of \textit{G-Nuggets} are available online at : \textit{\url{https://phoebs88.github.io/A-Catalogue-of-Game-Specific-Software-Nuggets}}.
    \item A dataset consisting of total 1185 text records, with 892 \textit{commits}, 189 \textit{issues}, and 104 \textit{pull requests} mined from 100 open-source games from GitHub available on the above mentioned website.
\end{itemize}

The remainder of the paper is structured as follows. Section \ref{methodology} focuses on the methodology used to obtain the catalogue. Detailed catalogue of \textit{G-Nuggets} is presented in Section \ref{catalogue} with discussion in Section \ref{discussion}. Threats to validity are discussed in Section \ref{threats} followed  by related work in Section \ref{related}, and eventually the paper ends with conclusion and future work in Section \ref{conclusion}.

\section{Methodology}
\label{methodology}
% Existing literature shows that researchers have used number of techniques to detect \textit{Software Nuggets} in a dataset such as machine learning based \cite{maiga2012support}, rule/heuristic-based \cite{moha2009decor}, metric-based approaches \cite{chidamber1994metrics}, and so on \cite{sharma2018survey}. Although these methods are significant enough, there are several shortcomings. For example, a metric-based approach is best fit for anti-pattern cataloguing where we can apply certain metrics with their threshold values. However, it lags in providing desired results in the case of flexible data with large variation. Datasets can be subjective, and thus metric-based approach may not be efficient. Another example can be of the rule-based detection method. A rule-based anti-pattern detection is suitable in the context where the data is clear, precise, and unambiguous, but it lags in the indeterminate dataset as it may contain \textit{Software Nuggets} in multiple forms which cannot be captured by rules alone. Thus, considering these factors, we choose to apply a methodology that can take advantage of multiple approaches. We resort to combine a few of the existing methods to propose an ensemble of methods for obtaining game-specific \textit{Software Nuggets}. Hence, 

\begin{figure*}
\centering
    \includegraphics[width=18cm,height=4.5cm]{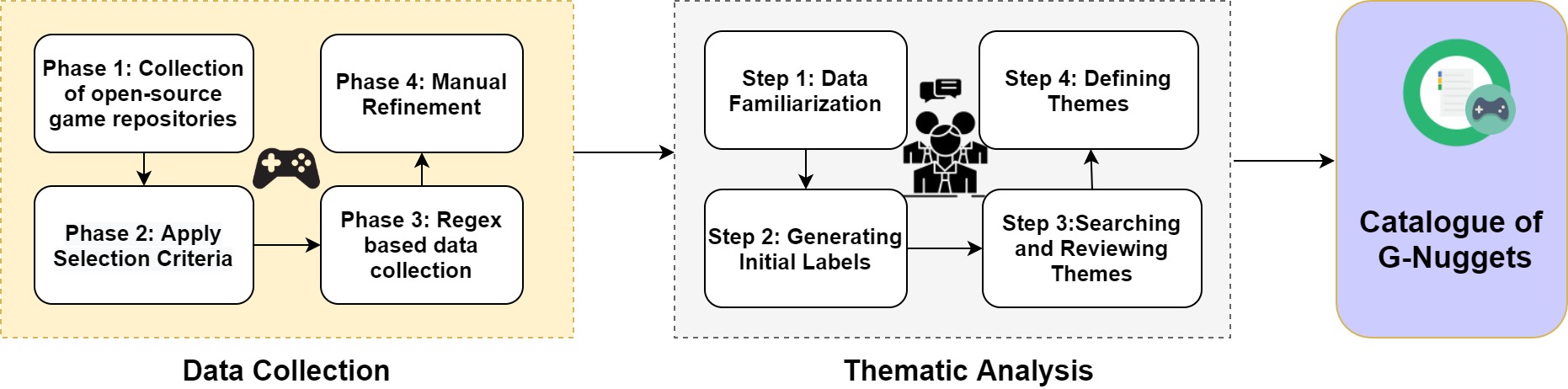}
    \caption{Methodology used to obtain catalogue of game-specific \textit{Software Nuggets}}
    \label{fig:approach}
\end{figure*}

We propose a four-phase data collection process followed by thematic analysis for categorization of \textit{G-Nuggets}. Thematic Analysis is a commonly used methodology used by researchers to analyze patterns by utilizing qualitative analysis \cite{fereday2006demonstrating}. Researchers have used this qualitative method for categorizing energy patterns \cite{cruz2019catalog}, agile challenges \cite{gregory2015agile}, awareness interpretation for collaborative computer games \cite{teruel2016applying}, and so on. Researchers also performed thematic analysis to gain insights based on user reviews for disaster apps \cite{tan2020modified}. Thus, leveraging the thematic analysis process fits our context of catalogue creation. We collect the dataset from 100 open-source GitHub game repositories of different genres, such as board game, puzzle, arcade, and so on. Methodology is shown in Figure \ref{fig:approach}, which comprises of two tasks:
\begin{itemize}
    \item Collection of the dataset in the form of \textit{commits}, \textit{issues}, and \textit{pull requests}, from open-source games mined from GitHub.
    \item Execution of thematic analysis on the collected data to obtain \textit{G-Nuggets}.
    %\item Implementation of Latent Dirichlet Allocation (LDA) over whole dataset for validation and refinement of categories.
\end{itemize}

\subsection{Dataset Collection}
\label{datacollection}
%phase 1: where we collected links of game info of 229 games
%phase 2: filtered games on selection criteria and ended up with 100 games
%phase 3: regex search
%phase 4: manual filtering of unnecessary text records
%then applied thematic analysis

We followed a multi-phase process to collect the dataset of text records for thematic analysis in the process of generating the catalogue for \textit{Software Nuggets}. We describe them below:

\subsubsection{\textbf{Phase 1: Collection of open-source game repositories}}
We started collecting the game repositories from a popular GitHub repository\footnote{\label{note1}\url{https://github.com/leereilly/games}} that provides awesome list of desktop games. The GitHub repository\textsuperscript{\ref{note1}} listing the popular games itself has the stargazers count of approximately 17.6K. 
We gathered the repositories and browsed each game to inspect if it is game by checking its README as there were non-game repositories and forked repositories of the original source.
Thus, we ended up with a list of URLs of 229 game repositories. There were a total of 15 genres of games spread across 19 programming languages. The three major languages in which these games were developed are \textit{JavaScript}, \textit{C++}, and \textit{C} with 92, 45, and 23 games respectively. The games present in this list were either \textit{browser-based} (can be played instantaneously in the browser with no need for installation) or \textit{native}, which requires installation.

\subsubsection{\textbf{Phase 2: Apply selection criteria}}
Although we collected links of 229 games, we observed that there were games with zero star count, zero forks, and with size of less than 1 MB. Therefore, for thematic analysis we decided to resort to a subset of these games repositories. We proposed the selection criteria below and filtered 100 games out of 229 game repositories. We took the subset of the total games, which is significant enough and can represent the whole dataset as elaborated in below description. We created the subset by combining random games and the games having more stargazers count. Thus, mathematically subset can be shown as:\newline

\fbox{\begin{minipage}{23em}

\textbf{Selection Rule}\newline
If \textit{A = Whole Dataset}, then $B\subset$A, \newline
  where, \textit{B = 0.5*(top starred games(A))+0.5*(random sample of remaining dataset after selecting top starred games(A))}
 
\end{minipage}}

\begin{itemize}
\item \textbf{First Half :} Half of the subset determined for thematic analysis contains the top games sorted according to the stargazers count in descending order. We did this to make sure that we include the games which are most popular among the developer community \cite{borges2018s,ray2014large}. Also, the games having more stargazers count are generally open-source games which are not developed by a single developer, but a team of them \cite{klug2016understanding}. Thus, we consider and analyze these games for our catalogue. We selected a total of 50 top starred games from the 229 game repositories. The metadata of the half subset is mentioned below:
    \begin{itemize}
        \item Average stargazers count: 2320.9
        \item Average forks count: 816.46
        \item 22 games out of 50 are developed in the \textit{C++} language.
    \end{itemize}

  \item \textbf{Second Half :} Another half of the subset belongs to the random sample taken from the remaining links after the selection of top starred games. We did so to ensure that we also take the randomized sample of the whole dataset.
To confirm that the random sample is statistically significant, we resort to measuring the confidence level and confidence interval of the selected subset. The confidence interval represents the range of data that incorporates the true value of the unknown population parameter. In other words, if we construct an infinite number of the independent sample using a confidence interval, then the confidence level will correspond to the proportion that contains the true value of the parameter. Therefore, based on stargazers count as a metric, the confidence level of the selected subset is 95\% with a confidence interval in the range of 70-135 \cite{kitchenham2002preliminary}. A confidence interval of 95\% states that if we consider random samples of the same sample size for 100 times, then 95 out of 100 contains the true but unknown mean in the interval of 70-135. A total of 50 games were selected randomly from the set of 179 games (deducting top starred 50 games from 229 games). The metadata of random games selected is mentioned below.
    \begin{itemize}
        \item Average stargazers count: 102.5
        \item Average forks count: 43.2
        \item 35 games out of 50 were developed in \textit{JavaScript} language.
    \end{itemize}
\end{itemize}
We observe that selected games in the subset majorly written in \textit{C++} and \textit{JavaScript}. Further, among the 100 games selected, 44 games are \textit{native} games while the remaining 56 games are \textit{browser-based}. The metadata of the game chosen for the study is shown in Table \ref{meta}.

\subsubsection{\textbf{Phase 3: Regular expression based gathering of data:}} Based on the 100 games selected for analysis, we made a regular expression based search in \textit{commits}, \textit{issues}, and \textit{pull requests} to find out the subjects of our potential interests \cite{cruz2019catalog, bao2016android}. We searched for various words that can possibly correspond to \textit{G-Nuggets}. The search words along with the rationale for selecting those words are given in Table \ref{regex}. We include few words such as:
\begin{lstlisting}
.*(performance|efficiency|delay|lag|
usability|refactor|code smell|anti-
pattern|bad|issue|bug|defect|flaw|
fault|problem|energy|battery|power|
freeze|crash|hang|glitch|control).*
\end{lstlisting}

\begin{table}[h]
\centering
\caption{Rationale behind words chosen for Data Collection }
\label{regex}
\begin{tabular}{|l|l|}
\hline
\textbf{Terms}& \textbf{Rationale}\\
\hline
Performance, Efficiency, & To trace the records having issues\\
Delay, Lag&related to performance of games. \\
\hline
Usability & To seek the problem related to\\
&usability.\\
\hline
Refactor, Code Smell,  & Problem related to bad practices \\
Anti-Pattern&in code and their refactoring.\\
\hline
Bad, Issue, Bug, Defect, & To track the problems which\\
Flaw, Fault, Problem&causes some issue during the game \\
&play. We made it general to\\
&track any kind of issue in games.\\
\hline
Energy, Battery, Power& Problems related to energy\\
&efficiency of game.\\
\hline
Freeze, Crash, Hang,& To track the issues related \\
Glitch&to the user interface of game.\\
\hline
Control& Issues related to the control system \\
&of games.\\
\hline
\end{tabular}
\end{table}

We used GitHub API v3\footnote{\url{https://developer.github.com/v3}} and PyGithub\footnote{\url{https://pygithub.readthedocs.io}} to mine \textit{commits}, \textit{pull requests}, and \textit{issues}. For \textit{pull requests} and \textit{issues}, GitHub API v3 documentation\footnote{\url{https://developer.github.com/v3/issues}} states that -\textit{GitHub's REST API v3 considers every \textit{pull request} an issue, but not every \textit{issue} is a pull request}. For this reason, we identified \textit{issues} out of \textit{get\_issues (state=`all')} function, which returns \textit{issues} as well as \textit{pull requests}. Another function \textit{get\_pulls (state=`all')} returns a list of all \textit{pull requests}. We used result of both functions to segregate \textit{issues} and \textit{pull requests}. We merged the text data of the title, body, and comments of all \textit{issues} and \textit{pull requests} to do the regular expression matching. For \textit{commits}, we included text records, which are merged in the default branch of the GitHub repository. So, we mined a total of 1989 records from 100 games, which consists of 1245 \textit{commits}, 523 \textit{issues}, and 221 \textit{pull requests}.
We can observe that count of \textit{commits} is higher than \textit{issues} or \textit{pull requests}. 
The possible reason for this is the number of \textit{commits} in comparison to \textit{pull requests} and \textit{issues} are generally more in most of the repositories. Some of the repositories have no \textit{issues} or \textit{pull requests}; still, they have a lot of \textit{commits}. For example, Game-off-2013\footnote{\url{https://github.com/redbluegames/game-off-2013}}, is popular among the developer community with 588 forks so far and contains 368 \textit{commits} but no \textit{issues} or \textit{pull requests}. 

\subsubsection{\textbf{Phase 4: Manual refinement}}:
To validate that the subjects we mined are relevant and fit our interest, we manually inspected the data collected to separate the false positives \cite{cruz2019catalog}. One instance of false positive we got a comment in one of the \textit{issues} mined as \textit{We had a rocking day for gameplay balancing. I think it's pretty good for now, let's create separate issues for any remaining tweaks }\footnote{\url{https://github.com/lostdecade/onslaught\_arena/issues/9}}. We observe that the text data does not specify any anti-patterns in games.

We followed two strategies for manual refinement of the records:\newline
1. Examine the matched text data with the regular expression and find its relevance in the context of games.\newline
2. Analyze the entire thread of the matched record by visiting its GitHub page, and observing the comments and meaning of the conversation. It can help in removing the records that do not discuss the problems in games, but something else which is not consequential enough. For example,
we found an issue, where the developers were considering the updates and contribution to the repository and not about the problems in game\footnote{\url{https://github.com/KeenSoftwareHouse/SpaceEngineers/issues/584}}.

Thus after manual refinement of all records, we end up with 1185 text records with  892 \textit{commits},  189 \textit{issues}, and 104 \textit{pull requests}.

 \textbf{Data format}: Our dataset contains fields such as \textit{username}, \textit{name of the game}, \textit{URL}, \textit{text} (contains commit\_message, if the record is of the commit, otherwise contains the body of \textit{issues} or \textit{pull requests}), matching text with regular expressions, and a column named \textit{full\_content} which contains the combined text data of head, body, and comments in case of \textit{issues} or \textit{pull requests}. We made the column \textit{full\_content} to analyze the full-text data related to any \textit{issues} and \textit{pull requests}.
 
% We collected dataset based on the popularity of open-source games on GitHub. \cite{borges2018s} state that stargazers count of a GitHub repository represents its popularity among the developer community. It shows the count of people interested in a particular open-source project. Thus, considering the stargazers count, we collected a total of 229 games from a curated list of popular desktop games   \footnote{\label{note1}\url{https://github.com/leereilly/games}} and mined text data of \textit{commits}, \textit{issues}, and \textit{pull requests}. The GitHub repository\textsuperscript{\ref{note1}} listing the popular games itself has the stargazers count of approximately 14.2k. Below we discuss the process in detail.

% change the argument that we considered open -source games 

\begin{table}[h]
\centering
\caption{Metadata of 100 selected games}
\label{meta}
\begin{tabular}{|c|c|}
\hline
\textbf{Features}& \textbf{Statistics}\\
\hline
\hline
Average stargazers count& 1214\\
\hline
Average forks count& 431\\
\hline
Open issues& 92 \\
\hline
\#genres & 13\\
\hline
\#programming languages & 15\\
\hline
\# browser-based games& 56\\
\hline
\# native games&44\\
\hline
\end{tabular}
\end{table}
% \begin{figure}
%     \includegraphics[height=6cm, width=\columnwidth]{native_games.png}
%     \caption{Distribution of Native Games across various genres}
%     \label{fig:native}
% \end{figure}

% \begin{figure}
%     \includegraphics[height=6cm, width=\columnwidth]{browser_games.png}
%     \caption{Distribution of Browser-based Games across various genres}
%     \label{fig:browser}
% \end{figure}

%------------------------

%--------------------

\subsection{Thematic Analysis}
\label{thematicanalysis}
To curate the catalogue, we resort to the commonly used thematic analysis approach \cite{fereday2006demonstrating} and identify \textit{G-Nuggets} gathered from \textit{issues}, \textit{commits}, and \textit{pull requests}. In total, two researchers and one volunteer were involved in the whole process of thematic analysis. We followed the approach of thematic analysis by implementing below four steps on our dataset:
\begin{itemize}
\item \textbf{\textit{Data Familiarization:}} We thoroughly analyzed each record of our dataset and inspected the text data related to \textit{issues}, \textit{commits}, and \textit{pull requests}. We observed the title, body, and comments (column full\_content) of all records and discussed it with co-author and a fellow volunteer.

\item \textbf{\textit{Generating Initial Labels:}} 
Based on records of \textit{commits}, \textit{issues}, and \textit{pull requests}, we started giving initial codes to the dataset. We initialized some themes such as usability, performance, and many more on an abstract note. We divide the process into several iterations supported by rigorous discussions among both the authors and a fellow volunteer.

\item \textbf{\textit{Searching and Reviewing Themes:}} After analyzing all the instances of our dataset, we discuss and review them to find the relevance of the themes in the context of games. We divided them into subcategories wherever required and also merged the themes accordingly. We decide to discard the themes which occurred for few instances, i.e., less than three times. %All these themes are considered under the umbrella of \textit{G-Nuggets}.
% were very few and didn't occur for more than three times, we discarded them.

\item \textbf{\textit{Defining Themes:}} In this stage, we made an orderly description of each \textit{G-Nugget} and its occurrence. We name each theme as a \textit{G-Nugget}. Section \ref{catalogue} describes all themes observed for game-specific \textit{Software Nuggets} along with their definition and the occurrences in the dataset.  
% consists of all the themes we define for game-specific \textit{Software Nuggets}.
% It also contains the definition of each one of them with the number of occurrences in the 1185 records and their example.

\end{itemize}
Both the researchers and one volunteer were involved in all the stages of thematic analysis. There also arose times when the researchers were in disagreement on the themes, but then we analyzed the theme in the context of games, how the theme can help, and how it can affect the game development \cite{shi2015game}. Also, there were instances where a single record was falling into multiple themes. The co-occurrences can be found here\footnote{\url{https://osf.io/jqv9m/?view_only=ab7a669e6faf41798a9e6212b2485c6e}} (Please download the file for better view).

In total, 892 \textit{commits}, 189 \textit{issues}, and 104 \textit{pull requests} were analyzed during the process of thematic analysis. As a result, we obtain ten game-specific \textit{Software Nuggets}.

\section{Game-Specific \textit{Software Nuggets}: Catalogue Definition}
\label{catalogue}
Here, we list all the game-specific \textit{Software Nuggets}. We describe each \textit{G-Nugget} with the following: \textit{context, problem, solution, example} situation illustrating the occurrence in the case of games and \textit{implication}. For each \textit{G-Nugget}, we discuss implications for researchers (indicated with the symbol
 \textbf{\textit{R}}) and/or practitioners/developers (\textbf{\textit{P/D}}) based on our findings. A detailed discussion on each \textit{G-Nugget} along with the GitHub link of its occurrences is available online at: \textit{\url{https://phoebs88.github.io/A-Catalogue-of-Game-Specific-Software-Nuggets}}. Further, Table \ref{fig:example} shows examples of each theme along with its reference.
\begin{enumerate}
\item \textbf{Beware of Vague Inheritance from Game Engines/Frameworks}\\
% Games are often developed by building on top of existing game engines.  But the usage of game engines can sometimes cause the anti-patterns to get inherited in the game. \\
\textbf{Context}: Nowadays, games are often developed using a framework or game engine that simplifies the task of game developers by providing a set of templates/ modules which prevent them from developing the game from scratch \cite{lewis2002game}.\newline
\textbf{Problem}: The anti-patterns inherited can affect functional and non-functional requirements of the game.\newline
\textbf{Solution}: We should be aware of the anti-pattern present in the game engine/frameworks and should resolve it before using it for the development of the game. Game engines should be examined for the existence of any anti-pattern which can be inherited by the game, and thus should be removed at the time of development.\newline 
\textbf{Example}: Consider a board game using a chess library, that is used for chess piece placement/movement, move generation/validation, and check/checkmate/stalemate detection. Thus, if there exists an issue in any one of the rules of checkmate detection, then it can lead to the wrong detection for the game, that gets inherited from the chess library.\newline
\textbf{Implications}: \textbf{\textbf{R}} can explore this domain more in the context of games and how vague inheritance of \textit{Software Nuggets} could lead to unwanted consequences in game quality. \textit{\textbf{P/D}} should keep track of the \textit{Software Nuggets} present in the development environment to avoid unexpected technical debt in the game.

\item \textbf{Unwanted Movement of Game Objects}\newline
% Objects or icons which are making an unwanted move in games and vice versa affects the usability of game.\newline
\textbf{Context}: Player moves are one of the important parts of gameplay. Player movements should be logical, consistent and optimized \cite{korhonen2006playability}.\newline
\textbf{Problem}: If the game does not follow the expected behavior, it causes hindrance in the usability of the game. It does not cause any violation in the game rules, but it makes the game clumsy to play.\newline
\textbf{Solution}: Developers should keenly focus on this \textit{G-Nugget} by planning out the object movements in advance which is logical and expected by the end-user. \newline
\textbf{Example}: Consider the case of a board game such as chess, which requires grabbing and moving the pieces in the game. Thus, the player should not be allowed to move jail pieces (as shown in Figure \ref{fig:unwanted}). Similarly, there should be a check on moving pieces so that the players do not move the piece, which is not of his/her color. It avoids unnecessary extra work and needless confusion for end-users.\newline
\textbf{Implications}: For researchers \textit{\textbf{R}}, finding out the frequent occurrences of this \textit{G-Nugget} and the steps leading to them can be a interesting research direction in the context of game. For developers \textit{\textbf{P/D}}, this \textit{G-Nugget} can help in optimizing the game object movements, thus leading to better performance of game.
 \begin{figure}
    \centering
    \includegraphics[width=\columnwidth, height=4cm]{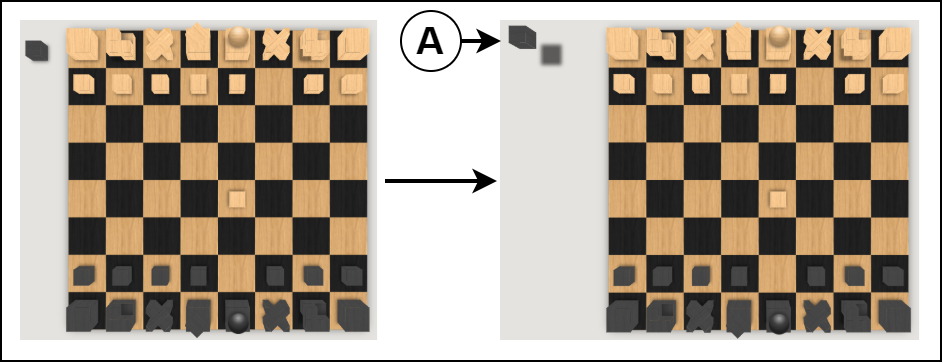}
    \caption{A scenario of \textit{Unwanted moves} taken from open-source game, \href{https://github.com/juliangarnier/3D-Hartwig-chess-set}{\underline{\textit{3D Hartwig Chess Set}}}. Label [A] shows the unwanted movement of jail pieces.}
    \label{fig:unwanted}
\end{figure}

\item \textbf{Minimum Game Controls with Maximum Game Functionality}\newline
% Game control actions involving mouse, keyboard, joystick, and other devices, should be smooth, and the actions should be optimized as it increases the usability of the game \cite{korhonen2006playability}.\newline 
\textbf{Context}: Game control is an integral part of the gameplay. Having effective, smooth and minimized movements of various devices such as keyboard, joystick, mouse, etc. increases the usability of game \cite{korhonen2006playability}.\newline
\textbf{Problem}: Rough and faulty game controls can have a negative impact on usability factor.\newline
\textbf{Solution}: Optimized game actions along with smooth game controls increase hassle-free gameplay.\newline
\textbf{Example}: In the case of point and click game, there should not be any action related to the keyboard or any other device, as it causes unnecessary control movements for the player.\newline
\textbf{Implications}: Researchers \textit{\textbf{R}} can explore this \textit{G-Nugget} by proposing different ways to optimize game controls with maximum game features coverage. \textit{\textbf{P/D}} should focus on this \textit{G-Nugget} and provide game with minimum mouse scrolls, keyboard buttons, and other devices.

% \begin{figure}
%     \centering
%     \includegraphics[width=8cm, height=4cm]{confirmation.png}
%     \caption{Confirmation Dialog box affects usability of game.}
%     \label{fig:confirm}
% \end{figure}
    \begin{figure}
    \centering
    \includegraphics[width=\columnwidth, height=6cm]{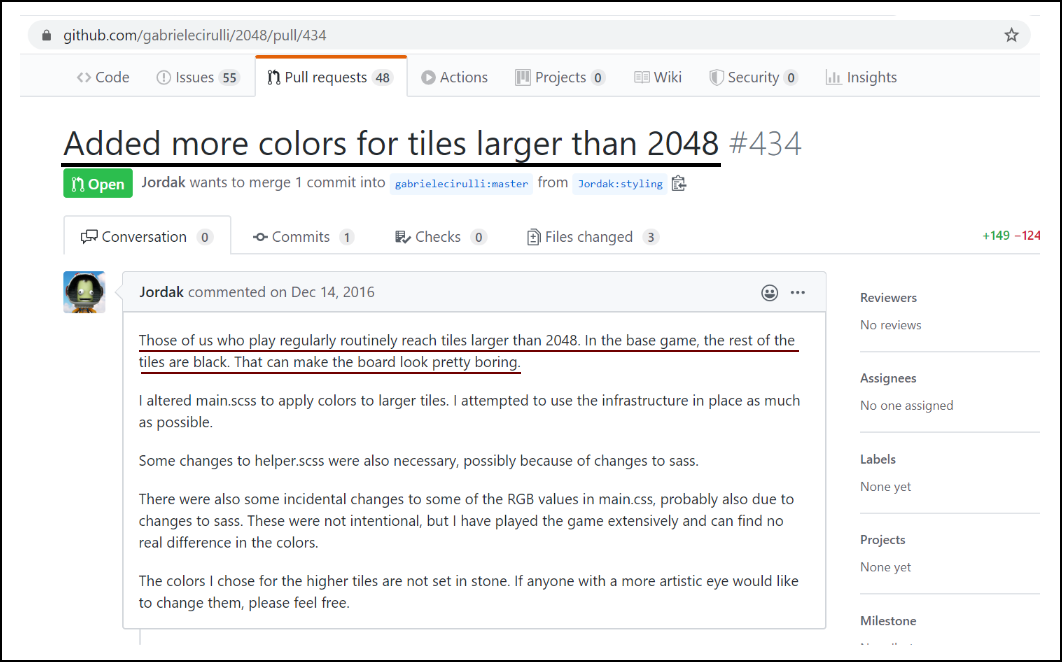}
    \caption{The screenshot is taken from a \textit{pull request} of the game \href{https://github.com/gabrielecirulli/2048/pull/434}{\underline{\textit{2048}}}, where a user talks about more colors for tiles as less colors make game boring. Underlined sentences highlights the text discussing about the UI Design of game.}
    \label{fig:uidesign}
\end{figure}
% \item\textit{\textbf{Confirmation Box}}\newline
% We also include the issue of users complaining about no confirmation box before any decision. Absence of confirmation box before any important decision of game such as restart of the game, jumping to the next level, and quitting the game, affects the usability of the game. \newline
% \textbf{Context}: Players are often required to switch from one scene to another during the gameplay. The switching of scenes should be according to the wish of the user. \newline
% \textbf{Problem}: Unnecessary change of scene without confirmation from the player causes disinterest and discomfort in user.\newline
% \textbf{Solution}: To avoid above situation, there should be a confirmation box before any important decision in the gameplay.\newline
% \textbf{Example}: Sometimes, players hit the wrong button, say exit game. In case if the game does not have a confirmation box for the same, the game would end. This kind of scenario leads to the disinterest of players in the game (as shown in Figure \ref{fig:confirm}). 
% %\end{enumerate}
\begin{figure}
    \centering
    \includegraphics[width=\columnwidth, height=6cm]{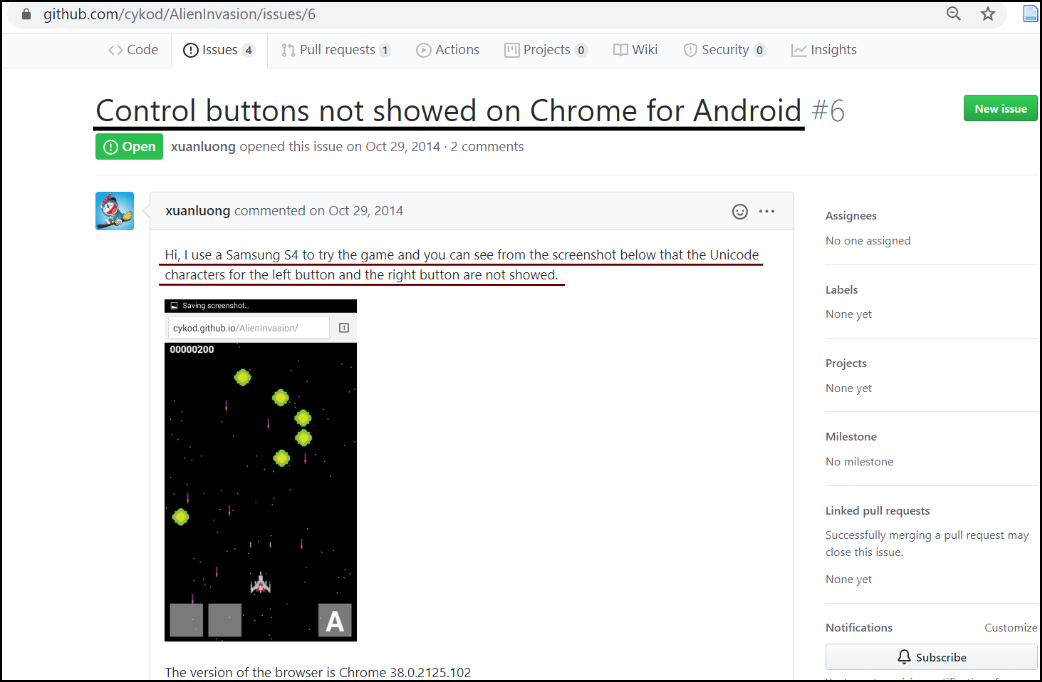}
    \caption{Screenshot taken from one of the \textit{issue} of game \href{https://github.com/cykod/AlienInvasion/issues/6}{\underline{\textit{AlienInvansion}}}, where the user is not able to view the game controls properly on a particular Chrome version and specific mobile phone.}
    \label{fig:builddiag}
\end{figure}
% \item\textbf{Design}\newline
% Design of game is one of the leading factors which decides the future of the game. We divide this category into three sub-themes:
% \begin{enumerate}
\item \textbf{Avoid Wrong Logic/Invalid Moves}\newline
% Game design involves the implementation of many small modules that results in one big module. These small modules of functionalities should be implemented properly with proper logic; otherwise, it can lead to a faulty game. Invalid moves lead to a violation of game rules and thus creates confusion for users. \newline
 \textbf{Context}: Game design involves the implementation of many small modules that results in one big module. These small modules of functionalities should be implemented properly with proper logic; otherwise, it can lead to a faulty game. Invalid moves lead to a violation of game rules and thus creates confusion for users.\newline
\textbf{Problem}: Violation of game rules through invalid moves can destroy the likeness of the game in player's mind as it highlights the loopholes in a game. \newline
\textbf{Solution}: Wrong Logic/Invalid moves should be strictly limited by the developers by carefully defining the control-action of the game. There should be precise ``Game Description Language'' which can guide the developer on each step \cite{thielscher2010general}.\newline
\textbf{Example}: In a chess game, if the game design allows illegal moves also in the game, without any objection, then the game may lose its purpose of play.\newline
\textbf{Implications}: This \textit{G-Nugget} opens up new research direction for researchers \textit{\textbf{R}} to explore it deeply and how it affects the gameplay. Further, open-source game developers \textit{\textbf{P/D}} should particularly try to avoid and check invalid moves in their game.

\item \textbf{User Interface Glitch}
%  The UI related issues which leads to bad UI of the game are included in this theme.\newline
\textbf{Context}: UI is an essential component of game display that can attract or bore the user.\newline
\textbf{Problem}: Bad UI design can lead to a displeasing display of the game. Issues which cause problems in the audio/video of games affects the UI of game and can lead to demotivation in the player for playing the game. \newline
\textbf{Solution}: The UI design should be visually pleasing and should follow the UI heuristics of game \cite{johnson2003effective}.\newline
\textbf{Example}: The tile size in the board related games should be of average size. It should not be too big or too small. Another instance can be shown in the screenshot (Figure \ref{fig:uidesign}) that discusses tiles of more colors in the game 2048.\newline
\textbf{Implications}: Open-source developers \textit{\textbf{P/D}} should specially emphasize on UI of game by conducting UI quality assurance test. They should test the compatibility of game with various devices and should ensure proper layout, content, images and font.

\item\textbf{Be Cautious of Platform Dependency}\newline
% Platform Dependency is one of the critical and initial aspects before game execution, which cannot be ignored. \newline
\textbf{Context}: Games come with a set of dependency on how they have to run. Different platforms and versions are required to execute the game. Often the installation process of a game takes more time than expected due to the version problem, error in Makefile, and other reasons.\newline
\textbf{Problem}: If the platform dependencies are not stated explicitly, it may cause problems for user. Faulty installation can lead to disinterest of user from the beginning only.\newline
\textbf{Solution}: Documentation of the game dependencies must be provided to end-users to ensure hassle-free game play.\newline
\textbf{Example}: Consider a game running smoothly on a Windows laptop, but not on Mac PC. Further, the developers did not provide any information about the same. Another example can be related to the processor requirement of games. All the requirements and dependencies should be documented beforehand so that it does not effect the gaming experience of a user. Figure \ref{fig:builddiag} shows a screenshot of an \textit{issue} related to platform dependency of game, where the user faces problem in viewing the icons and game controls because of platform dependency.\newline
\textbf{Implications}: All the steps and requirements related to platform dependency of the game should be properly documented involving lower-level details by the developer \textit{\textbf{P/D}}; thus the user will not have to struggle on the execution of game.

% \item \textit{\textbf{Installation}}\newline
% Installation is the step we do after determining the dependency of any game.
% \textbf{Context}: Often the installation process of a game takes more time than expected due to the version problem, error in Makefile, and other reasons.\newline
% \textbf{Problem}: Since it is the preliminary requirement for execution of any game, faulty installation can lead to disinterest of user from the beginning only.\newline
% \textbf{Solution}: The installation process should be smooth, user-friendly, and properly documented. Developers should check for the mistakes in the installation process in advance, to avoid unnecessary disinterest of end-users. \newline
% \textbf{Example}: Build issues related to Makefile of source code can cause installation problems.
% \end{enumerate}
%%%%%%%%%%%%%%%%from here
\begin{figure*}
    \centering
    \includegraphics[width=15cm, height=5cm]{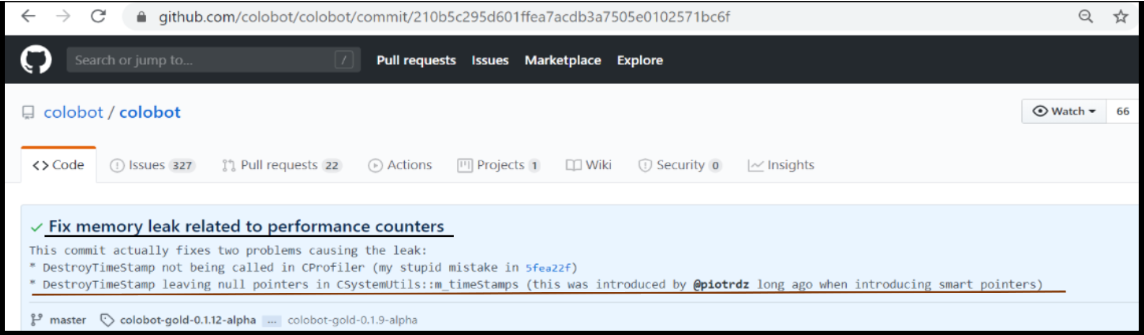}
    \caption{Screenshot of a \textit{commit} done by developer to rectify the memory leak problem caused by usage of null pointer in game \href{https://github.com/colobot/colobot/commit/210b5c295d601ffea7acdb3a7505e0102571bc6f}{\underline{\textit{Colobot}}}.}
    \label{fig:memorydiag}
\end{figure*}
% \item \textbf{Performance}\newline
% Anti-patterns affecting the performance of games are categorized in this theme.\newline
% \begin{enumerate}
% \item \textit{\textbf{Lag/Delay}}\newline
% Poor performance, such as unexpected delay, lag, hindrance in the flow of the game, and unacceptable game environment, leads to the disinterest of a player \cite{dick2005analysis}.
% \newline
% \textbf{Context}: Often players face difficulty during the game such as game lag during multiplayer network-based gameplay, delay in an animation of the game, and so on which causes poor performance of game.\newline
% \textbf{Problem}: Poor performance of game affects the usability of the game and can annoy the user.\newline
% \textbf{Solution}: The developer should pay special attention to the factors that lead to the poor performance of games such as game animation, switching of scenes, network-based gameplay \cite{dick2005analysis}. Reason of lag/delay can also be due to insufficient specification of system required for the gameplay.\newline
% \textbf{Example}: Consider a high-end game played on a low specification computer.

\item \textbf{Avoid Memory Leaks}\newline
% Memory leak that affect the performance of games is categorized in this theme.\newline
\textbf{Context}: Games involve different game objects and assets, which consumes a lot of memory. This memory storage should be used wisely; otherwise it can cause memory leak.\newline
\textbf{Problem}: Memory leak causes reduction in available memory for usage, which causes bad performance. \newline
\textbf{Solution}: Memory leak can be reduced by avoiding number of case scenario such as : multiple references to the same object, creating huge object tree, assurance of proper garbage collection of unused variable.
\textbf{Example}: Consider an arcade game, where different game objects keep on appearing and disappearing in the gameplay. These objects need to be handled carefully in the memory to avoid a memory leak. Figure \ref{fig:memorydiag} shows a screenshot of game where developer amend the code to remove the problem of memory leak in game.\newline
\textbf{Implications}:  To avoid memory leak problem, developers \textit{\textbf{P/D}} should free up the unnecessary variables and game objects, and should use the idea of data locality \cite{nystrom2014game}. Developers should use the object pool where memory can be reused instead of allocating and freeing them every time \cite{nystrom2014game}.

\item \textbf{Avoid Energy Extensive Patterns}\newline
% Sustainable software development has become one of the key requirement of current era. We considered the necessity of energy-efficiency in this theme. \newline
\textbf{Context}: Sustainable software development has become one of the key requirement of current era. Any game should be sustainable and energy-efficient which consumes optimized power \cite{capra2012software}. It must be designed in a way so that it fulfills the functional and non-functional requirement with minimum power usage. \newline
\textbf{Problem}: More power consumption leads to faster power drainage.\newline
\textbf{Solution}:Developers should avoid energy \textit{Software Nuggets} in game \cite{cruz2019catalog}, and should utilize the process of sustainable software development \cite{amsel2011toward}.\newline
\textbf{Example}: Consider a game where graphics and animation in games are active even after game over. This causes unnecessary power consumption.\newline
\textbf{Implications}: Optimization of energy has been focused by researchers \cite{cruz2019catalog, amsel2011toward}, however there is need to further explore energy consumption in the context of games more deeply. They \textit{\textbf{R}} can emphasize on the possible ways on how game can be more interactive with optimized power consumption. Further, practitioners and open-source developers \textit{\textbf{P/D}} should follow good practices to reduce the power usage of their of games.

\begin{table*}
  \centering
    \caption{Themes identified during Thematic Analysis (n = number of occurences in the dataset)}
    \label{fig:example}
        \begin{tabular}{|c|c|c|c|}
        \hline
        & \textbf{Themes} & \textbf{Example} & \textbf{Ref}\\
        \hline
1.&Beware of Vague Inheritance from Game &\textit{``Chess.js library returns a location string with only a number}&\href{https://github.com/juliangarnier/3D-Hartwig-chess-set/pull/9}{Link}\\
&Engines/Frameworks&\textit{and causes the player to be locked with the piece,not letting them}&\\
&(n=65)&\textit{ move it at all. This code checks if the string is the correct}& \\
&& \textit{length before allowing it in hideMoves and showMoves to prevent}&\\
&&\textit{ the error from being thrown ,as it runs just fine so}& \\
&&\textit{ long as that faulty object is excluded."}& \\
\hline
%----------------------------------
2.&Unwanted Movement of Game Objects &\textit{``Fix automovement toggling on ``joystick used" flag." }&\href{https://github.com/OpenMW/openmw/commit/ef2a7160fa21142fe0f4d6fdac2cafa40a651c08}{Link}\\
&(n=63)&\textit{}&\\
\hline
3.&Minimum Game Controls with &\textit{``When holding down a key, the keyboard}&\href{https://github.com/fogleman/Craft/issues/230}{Link}\\
&Maximum Game Functionality&\textit{buffer fills, causing the associated behavior}&\\ 
&(n=191)&\textit{to continue afterward for some time."}&\\
\hline
%-------------------------------
4.&Avoid Wrong Logic/Invalid Moves &\textit{``put everything inside a function also fix}&\href{https://github.com/kenrick95/c4/commit/631377d39ef89cd566d2c6188440cf0318dfdbea}{Link}\\
&(n=200)&\textit{bug when AI is making invalid move"}&\\
\hline
%------------------------------------

5.&User Interface Glitch &\textit{``Drag tool makes graphic glitchy. Whenever I use the frag tool, }&\href{https://github.com/lo-th/3d.city/issues/33}{Link}\\
&(n=403)&\textit{I think the game lags and the graphics becomes glitchy."}&\\

\hline
%-------------------------------------
6.& Be Cautious of Platform Dependency &\textit{``Hi, I use a Samsung S4 to try the game and you can see}&\href{https://github.com/cykod/AlienInvasion/issues/6}{Link}\\ 
&(n=252)&\textit{from the screenshot below that the Unicode characters for the}&\\   
&&\textit{left button and the right button are not showed."}&\\
\hline
%------------------------------------------

7.&Avoid Memory Leaks&\textit{``Looking at Chrome's task manager, memory for the tab grows}&\href{https://github.com/particle-clicker/particle-clicker/issues/42}{Link}\\
&(n=53)&\textit{by up to 180k per second. This is unacceptable for}&\\
&&\textit{an idle/incremental game."}&\\
\hline

%--------------------------------------
8.& Avoid Energy Extensive Components&\textit{``game over was executing even after game over - no need really.}&\href{https://github.com/rishabhp/pappu-pakia/commit/4283d761e9aa2c789785d47b6dcea6b0f368caf6}{Link}\\ 
&(n=7)&\textit{ lets save memory power"}&\\
\hline

%-------------------------------------
9.& Provide Offline-Support&\textit{``As mentioned in issue \#24, having a cache manifest will enable}&\href{https://github.com/gabrielecirulli/2048/pull/26}{Link}\\ 
&(n=6)&\textit{ the game to work offline, especially useful for mobile devices."}&\\
\hline
%----------------------
10.&Ensure Game Security&\textit{``Usernames with spaces fail to authenticate."}&\href{https://github.com/RigsOfRods/rigs-of-rods/issues/2487}{Link}\\ 
&(n=9)&&\\
%-----------------------------------
\hline
%----------------------
    \end{tabular}
  \label{manualstudy}
\end{table*}

\item \textbf{Provide Offline-Support}\newline
% This theme is about the browser-based game.\newline
\textbf{Context}: Many games these days avoid the hassle of downloading the game and then installing it. They make the game \textit{browser-based} so that the user can directly play the game without any prerequisite.\newline
\textbf{Problem}: Sometimes users get disconnected from the internet in the middle of the game because of which they may lose the game. \newline
\textbf{Solution}: Provide offline support for the game, so that people can save their game to the home screen, and play offline.\newline
\textbf{Example}: Consider a \textit{browser-based} \textit{2048} game; having offline support will help the users in continuing the game.\newline
\textbf{Implications}: \textit{Offline Support} is the functionality that should be considered by developers \textbf{\textit{P/D}} in the case of \textit{browser-based} games for better acceptance across players' community.
\item \textbf{Ensure Game Security}\newline
% Security is one of the essential factor that needs to be considered in the context of the game to protect a user's identity and his/her game achievements. We include issues related to security in this theme.\newline
\textbf{Context}: Security plays an important role in gameplay to protect a user's identity and his/her game achievements. It ensures that untrusted clients do not intrude into the game. It also ensures that players do not take unfair advantage in the game by doing wrong practices to win, such as cheat codes, compromised environment, and so on \cite{yan2002security}.  \newline
\textbf{Problem}: Untrusted players can become a threat to the security of the game.  Unethical means of achieving the target in gameplay may downgrade the game.\newline
\textbf{Solution}: Following all the heuristics related to security issues in games \cite{yan2002security} is key solutuion to this \textit{G-Nugget}.\newline
\textbf{Example}: Consider an RPG (Role Playing Game),  where the user makes improper usage of cheat codes to achieve the target.\newline
\textbf{Implications}: Developers \textit{\textbf{P/D}} should ensure that there is a proper channel of authentication for players that checks for insecure passwords, multiple logins, sniffed passwords, and so on. They should check for the loopholes of the game that can compromise the security of the game. \newline

% \item \textbf{Networking Issues}\newline
% Internet connection is an unavoidable factor in the context of online games such as multiplayer RPG games. Issues related to networking are categorized in this theme.  \newline
% \textbf{Context}: Online games depend on the server for proper connection. Also, multiplayer online games depend solely on the network.\newline
% \textbf{Problem}: Any glitch in the network can result in a bad experience for the player. \newline
% \textbf{Solution}: For networking issues, developers should not only ensure good server connection, but they should also provide troubleshooting steps required to resolve the issues so that users can also try from his end for resolving the issue. \newline
% \textbf{Example}: Unsuccessful client-server handshake in online games.
\end{enumerate}

\section{Discussion}
\label{discussion}
In this paper, we attempted to present multiple dimensions of software quality bugs such as anti-patterns, code smells, bad practices, etc. into a single umbrella called as \textit{Software Nuggets}. We believe that taking this integrated approach might help developers to avoid bad practices during software development.
The catalogue proposed in this paper caters to ten \textit{Software Nuggets} in games. It can help game developers to avoid bad practices during game development. Although some of the nuggets we discussed are available in the literature, but in the context of games, we present this catalogue with the intention of aggregating all \textit{Software Nuggets} at one place. 
While we understand that there are a few broad categories in our catalogue, we kept it wide to include them in catalogue as they are already discussed in detail in the literature. 

We believe that this catalogue can be used as a guide for game developers to avoid the listed \textit{Software Nuggets} during game development. Although this catalogue can be useful for all game developers, but it could be more beneficial for open-source game developers as they lack guidance about the practices they should avoid during the development of the game. Whereas, the paid and commercial games have development teams to focus on different modules of game development. Yet, we feel, economical and paid game developers can also benefit themselves by following the catalogue of \textit{G-Nuggets} and avoiding the mistakes by carefully taking appropriate action. Thus, researchers and developers can reach out to catalogue to refer \textit{Software Nuggets} during game development. Researchers can use the dataset as well as the \textit{G-Nuggets} for further empirical studies, and also can improve both of them as well. 

% %\section{Results}
% %\label{results}
% %We mined \textit{\textit{commits}}, \textit{pull requests}, and issues from popular game repositories and generated a catalogue of anti-patterns existing in the domain of games by following the process of thematic analysis and topic modeling. We observed that primarily 11 anti-patterns are generally faced by end-users/developers. We also formed subcategories among them. All the categories with their subcategories, definition and examples are discussed in Section \ref{catalogue} and Table \ref{fig:example}. The occurrences of each category, the scripts that were written to mine the dataset, and results are available at: \url{https://phoebs88.github.io/A-Catalogue-of-Game-Specific-Software-Nuggets}.

\section{Threats to validity}
\label{threats}
Here, we discuss the potential threats to our research study, and how far we can generalize the catalogue.
%\subsection{Internal Validity}
We performed regular expression based search to find the \textit{issues}, \textit{commits}, and \textit{pull requests} that may contain \textit{Software Nuggets}. Although it covers the majority of the text data, however, there is the possibility of missing some relevant text data. We considered the English language primarily, assuming it is the most commonly spoken language among developer communities. Also, while manual filtering, we tried our best to delete all cases of false positives; however, the possibility of having a few more false positives in our dataset cannot be denied. 

We mined the dataset from GitHub, believing that it is the preferred open-source platform for developer communities; still, it may be the case that we did not cover all possible \textit{G-Nuggets}. Likewise, we chose \textit{commits}, issues, and \textit{pull requests} from GitHub to analyze the commonly occurring \textit{G-Nuggets} faced by end-users and developers. Nonetheless, there is a possibility that other methodologies could have resulted in different \textit{G-Nuggets}. Moreover, we considered the default branch of repositories while dataset mining as other branches are not validated by the development team, and we are not confident about their quality. Thus, the discussions made in other branches of projects are not considered in the scope of this study. 

To the best of our efforts, we obtain the catalogue by in-depth analysis and believe that it is factual. But, since it involved manual interpretation also, it can be inaccurate and incomplete. Also, we have done thematic analysis on a dataset of 1185 records. Thus, there is a possibility that considering a more extensive dataset would result in some more themes.

%\subsection{External Validity}
We considered popular desktop games (\textit{native} and \textit{browser-based}) in our dataset. We focused on popular open-source games for study. However, commercial and paid games may have different \textit{Software Nuggets} than the one we got. We tried to cover different genres in various languages for open-source games, but there is a possibility of getting different set of \textit{G-Nuggets} with the increased dataset having more variation. We primarily analyzed desktop and browser games for this study. There is a wide scope of inclusion of games from other platforms as well, such as Android, iOS etc.
% \subsection{Construct Validity}
% \textcolor{blue}{remove LDA from validity}
% The primary construct validity concerns of the current research include the validity of topic modeling done using \textit{LDA}, which may not represent the entire game dataset as well as the themes that we arrived at during the thematic analysis. We tried to mitigate these concerns by creating the catalogue using an ensemble of methods of regular expressions, \textit{LDA}, and thematic analysis and tried to come up with a consistent set of categories of anti-patterns.
\section{Related Work}
\label{related}
Game development has gained a fair amount of attention from researchers since the past decade. Although games are a kind of software, nowadays they acquire distinctive discussion from the research point of view \cite{murphy2014cowboys, wesley2016innovation}. Possible reasons for the difference in game development with that to the traditional software development is in the context of various performance and real-time factors being involved, such as memory allocation and de-allocation issues, rendering process, and graphical components \cite{nystrom2014game, politowski2016old, kanode2009software}. As an instance, Murphy et al. \cite{murphy2014cowboys} conducted a survey study with the game and non-game developers to find the substantial differences between the two, and found that the video game developers require a more creative mind, good knowledge of maths, and performance tuning in comparison to non-game developers. Similarly, Pascarella et al. \cite{pascarella2018video} studied 60 open-source projects to differentiate between games and non-games and analyzed that project organization, developers skills, automated testing, code reuse, and many other aspects differ in both the domains. In a preliminary work, Khanve \cite{khanve2019existing} has shown that the code-specific bad practices or code smells in games can be different from those of code smells in other software by manually analyzing the violation of game programming patterns in eight \textit{JavaScript} games and concluded that games need to be handled separately in terms of code smells. In a study on urgent updates of games on Steam platform, Lin et al. \cite{lin2017studying} concluded that 
those games which use a frequent update strategy have higher ratio of 0-day updates (updates that are released on the same day), in comparison to the games that use traditional update strategy. Further, Lin et al. \cite{lin2019empirical} performed an empirical study on game reviews of 6224 games on Steam platform, and found that the game reviews are different from the mobile app reviews.They emphasized that the amount of time users play a game before posting their review is a unique characteristic and can play a major role in future research studies for games.

Considering the importance of the game development process over conventional software development, researchers made attempts to study the problems in the game industry \cite{politowski2021game, jacob2011issues}. Petrillo et al. \cite{petrillo2008houston} proposed four main categories for issues related to computer games development by analyzing game postmortems: \textit{Schedule Problems, Budget Problems, Quality Problems, Management Problems,} and \textit{Business related problems}. %In a study, \cite{zagal2013dark} introduced three broad categories of dark design patterns in games: \textit{Temporal, Monetary, Social Capital-Based}, and described the dark pattern as \say{\textit{A dark game design pattern is a pattern used intentionally by a game creator to cause negative experiences for players.}}.
Yan et al. \cite{yan2002security} emphasized the security concerns in games such as cheating practices used by players by means of exploiting the source code, lack of secrecy, lack of authentication, and many other factors. Along with the potential ways of cheating in games, they also suggested the detection strategy and prevention measure for the same. A dataset is created on utilizing postmortem reports by developers which contains the software problems faced by them \cite{politowski2020dataset} . In a recent study, Borrelli et al. proposed a set of game-specific bad smells in Unity Projects \cite{borrelli2020detecting}.

Along with the game-related problems, researchers also proposed solutions to a few of these problems. Varvaressos et al. \cite{varvaressos2017automated} proposed automated runtime bug finding for video games based on \textit{game loop}. They experimented on six real-world games and tracked their bugs based on the games' bug database. Researchers propose various heuristics and measures to analyze the usability and friendly user interface of games \cite{korhonen2006playability}. A study discuss the code attributes required to develop quality games and thus illustrate the architectural desired practices with the help of a game \cite{graham2006toward}. In a recent survey of participants of Global Game Jam (GGJ), a 48-hour hackathon, researchers study the effects of time pressure over the quality of games \cite{borg2019video}. They concluded that GGJ teams rely on an ad hoc approach to develop and teams share contextual similarities to software startups.

The common factors among all the above studies are about their dataset, based on which they proposed their research work. They did analysis based on the data of source code of projects, postmortem reports, survey data, and so on. But, to the best of our knowledge, we are not aware of any study done to investigate games by utilizing GitHub data such as \textit{commits}, \textit{pull requests}, and \textit{issues}. GitHub is a vast source of information that should be considered to gain a better understanding of problems faced by the end-users during gameplay. There exist a number of empirical studies on GitHub data to analyze the various fields of software engineering, such as release engineering \cite{joshi2019rapidrelease}, code quality \cite{ray2014large}, software evolution \cite{silva2016we}, and so on. Researchers also studied the emotions of developers on GitHub \cite{huq2020developer}, recommender for \textit{pull requests} \cite{yu2014reviewer}, whereabouts of forks in a GitHub repository \cite{jiang2017and}, and many more. Several studies analyze the users' data on GitHub to extract the patterns of social and technical behavior \cite{yu2014exploring,weicheng2013mining}. But, our work is unique based on the insights we get from our dataset, which has not been studied so far in the context of games. Existing literature consists of studies on code quality, code smells, and anti-patterns in different backgrounds, but in the context of games, it is overlooked.
There are numerous catalogues proposed in the background of anti-patterns but in different domains such as architecture, energy-efficiency, and so on \cite{dalla2020towards, carvalho2019empirical}. Cruz et al. \cite{cruz2019catalog} analyzed energy patterns for mobile applications and resulted in 22 design patterns related to energy efficiency. A catalogue is proposed on four architectural smells along with their example and impact on code quality \cite{garcia2009toward}. Carvalho et al. curated a catalogue of 20 code smells prevalent in the presentation layer of android apps followed by a detection tool \cite{carvalho2019empirical}. Thus, to the extent of our knowledge, we are not aware of any catalogue, or listings that categorize the bad practices in open-source games from the perspective of developers/users by utilizing the GitHub data.  

\section{Conclusion and Future Work}
\label{conclusion}
We proposed a catalogue of ten \textit{Software Nuggets} in the context of games (named \textit{G-Nuggets}). Dataset was created by mining \textit{commits}, \textit{issues}, and \textit{pull requests} from popular open-source games available on GitHub. We considered the text data of the repositories to analyze the problems faced by game developers/end-users. We followed the process of thematic analysis to obtain themes related to game specific \textit{Software Nuggets}.
% We discussed all the themes in detail in Section \ref{catalogue} and also on the website \textit{\url{https://phoebs88.github.io/A-Catalogue-of-Game-Specific-Software-Nuggets}}. 
Among all, we found the \textit{User Interface Glitch} followed by \textit{Be Cautious of Platform Dependency} to be most frequent \textit{G-Nuggets}. To the extent of our knowledge, this is the foundational and first catalogue proposed in the context of games based on the dataset of \textit{commits}, \textit{issues}, and \textit{pull requests} of GitHub open-source game repositories. Since the catalogue proposed is prevailing in the context of games, it can help game developers make informed decisions inconsiderate of platform dependency. 

We plan to extend this research work by reverse-engineering the patterns to code level and propose corresponding game-specific code smells to each of the \textit{G-Nugget}. We also plan to develop an automated way to detect these \textit{Software Nuggets} of the game by the support of the tool. %We wish to analyze the accuracy of the tool based on fields we selected for our study, that is, \textit{issues}, \textit{commits}, and \textit{pull requests} in order to determine whether the text data of these fields affect the accuracy of the tool or not. The reason behind this can be that the text data
We further plan to revisit this study using a dataset containing the source code of games. Also, we wish to extend the catalogue to a broader scope by including games of different genres, such as Android, iOS, and many more. Further, to measure the magnitude of this catalogue work, we intend to conduct an empirical study to analyze the benefits gained by game developers.

\section*{Acknowledgements}
We would like to thank Ashish Kumar for his contribution to the process of thematic analysis and illustrations of the game-specific \textit{Software Nuggets}.

\balance{}
\bibliography{games}
\bibliographystyle{IEEEtran}

\end{document}